# scientific reports

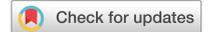

**OPEN**

# A self-contained and self-explanatory DNA storage system

Min Li[1,2,3], Jiashu Wu[1,2,3], Junbiao Dai[1], Qingshan Jiang[1], Qiang Qu[1], Xiaoluo Huang[1] & Yang Wang[1]✉

Current research on DNA storage usually focuses on the improvement of storage density by developing effective encoding and decoding schemes while lacking the consideration on the uncertainty in ultra-long-term data storage and retention. Consequently, the current DNA storage systems are often not self-contained, implying that they have to resort to external tools for the restoration of the stored DNA data. This may result in high risks in data loss since the required tools might not be available due to the high uncertainty in far future. To address this issue, we propose in this paper a self-contained DNA storage system that can bring self-explanatory to its stored data without relying on any external tool. To this end, we design a specific DNA file format whereby a separate storage scheme is developed to reduce the data redundancy while an effective indexing is designed for random read operations to the stored data file. We verified through experimental data that the proposed self-contained and self-explanatory method can not only get rid of the reliance on external tools for data restoration but also minimise the data redundancy brought about when the amount of data to be stored reaches a certain scale.

With the prevalence of big data-based applications, the massive quantities of data created across the globe each year are increasing in an exponential fashion[1], and it is expected that the global data storage demands will rapidly rise to 175ZB[2] by 2025, with 2.5 Exabytes per day, far exceeding the world's capacities that traditional storage technologies can afford[3].

On the other hand, given the uniqueness of gathered big data, it is often desired to archive them in external storage devices for value extending over a long period of time. To address these challenges, DNA is emerging as a novel storehouse[4–8] of information as it is not only potential to be orders of magnitude denser[9] than contemporary cutting-edge techniques but also extremely stable to retain the information for hundreds or even thousands of years[10], compared to hard drives which might last in only several decades[11,12].

Although it is a distinguished medium to store enormous data over millennia by virtue of its inherent high density[12–15] and durable preservation[16,17], DNA is still hardly practical as of today to store more than several hundred megabytes because of the high cost of DNA synthesis[18]. Consequently, most current research efforts are inspired to explore effective encoding and decoding schemes to facilitate the improvement of DNA storage density with reduced DNA synthesis cost as a goal[15,19–25]. For example, a common practical scheme is to compress a digital file before it can be transcoded into DNA sequence with added error correction[15,19,20,26,27] as the payload sent to the synthesis module so that the sequence payload can be instantiated into physical DNA molecules.

Despite that they are beneficial to storage density improvement, these schemes might result in high risks in data loss for ultra-long-term retention since the DNA storage systems are not self-contained anymore in the sense that the restoration of the stored DNA data has to rely on external tools, say *decompression program* in our example, which might not be available due to the high uncertainty in the far future. As a result, without solving this external reliance issue, it is unlikely for DNA storage to become a viable option for storing and retaining data in ultra-long terms.

In this work, to address the external reliance issue while improving the storage density, we propose a self-contained DNA storage system that can bring self-explanatory to its stored data without relying on any external tool. To this end, we allow the external tool (the corresponding decompression program) to be encoded with the compressed file together into a unified DNA sequence payload. However, unlike the one-to-one mapping between the compressed file and its corresponding decompression program in traditional cases, we deliberately make it possible to share a single decompression program among a set of compressed files for minimising the information redundancy. Although this strategy is not difficult to implement in traditional storage, it is hard to implement for the DNA storage given its random read hurdles and high sequencing cost during the data restoration. To address these issues and achieve our goal, we design a specific DNA file format whereby a *non-continuous* storage scheme is developed to reduce the data redundancy on the one hand, and combine with traditional storage media on the other hand to obtain effective indexing for random read operations to the stored data file while minimising the cost with an one-off sequencing.

[1]Shenzhen Institute of Advanced Technology, Chinese Academy of Sciences, Shenzhen 518055, China. [2]University of Chinese Academy of Sciences, Beijing 100049, China. [3]These authors contributed equally: Min Li and Jiashu Wu. ✉email: yang.wang1@siat.ac.cn

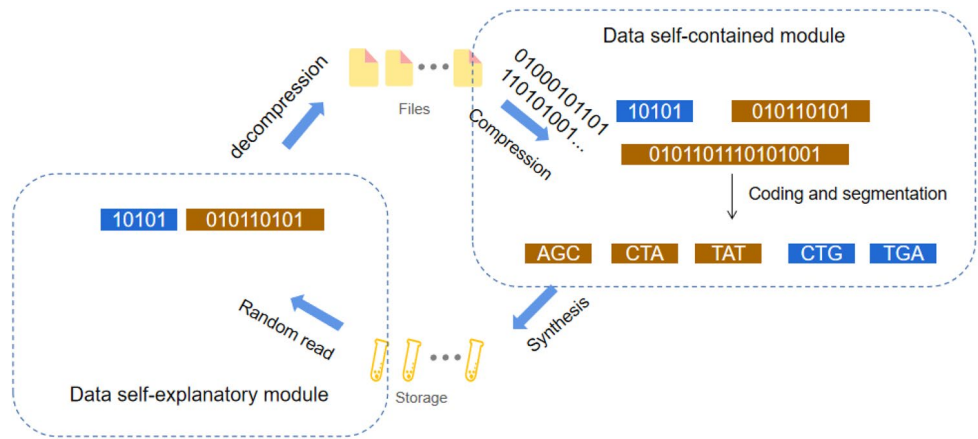

**Figure 1.** Overview of DNA storage process.

In summary, we made the following contributions in this paper:

1. For ultra-long-term data storage, we propose a self-contained and self-explanatory concept for DNA storage with an attempt to address the far-future uncertainties in DNA data restoration.
2. To realise the concept of the self-containment, we take data compression tool as an example to develop a DNA-based storage method that can not only minimise the cost of DNA synthesis and sequencing but also support random read operation from the DNA data.
3. To implement the storage method, we define a new file format to facilitate the self-explanation with read and write operations from/to the information stored inside the DNA file.
4. For the proof-of-concept, we apply 2 compression programs to 5 types of data files of different sizes to show the values and effectiveness of the proposed methods. The proposed methods achieved a 6–7% DNA nucleotide storage reduction, and the storage density reaches 90% of the ideal case.

Notably, our approach is novel in all of these aspects with a distinct aim to mitigate the challenges and risks faced by DNA data restoration in far future.

## Methods

In order to have a full play to the advantages that DNA can store data for a ultra-long time, we propose a concept of self-contained and self-explanatory technology for the DNA storage and design a method to implement it.

Since data compression is an important tool in the DNA storage for cost-efficiency, we concentrate in this research on the proposed technology by taking compression and self-extracting as a focus. In fact, the compression tool can also be stored with other data related information, such as encoding parameters, file storage format, etc. We describe our methodology in three steps. We first overview the DNA storage process in Fig. 1, and then introduce the detailed information regarding the data self-containment technology. Finally, we describe the data self-explanation technique by defining the format of the DNA file and DNA fragment to support the implementation of the functions presented in Fig. 2.

Figure 1 depicts the storage process where the input binary data file is often compressed to minimise the data redundancy while saving the synthesis cost. In order to achieve data self-containment, we store both the compressed data and the decompression program as the payloads in the DNA file. The binary data in brown represents the compressed data, and in blue represents the decompression program. Both types of data are segmented and uniformly coded as synthetic DNA sequence. Since it is necessary to distinguish between the data file and the program file in DNA fragments, we deliberately add a *bp-length* flag in the DNA fragments. The data self-explanatory process is reflected in the data restore process and is supported by the defined data format.

**Data self-containment.** The existing DNA storage systems are in general not self-contained as they always resort to external tools to backup/restore the data. For example, if the compressed data needs to be restored, the corresponding decompression program should be available. Although the chance of unavailability of the external tool is very small, we may still take the risk to lose the stored data in the case that the required tool is unavailable due to the uncertainties after a ultra-long time period for the DNA storage, say over hundred years.

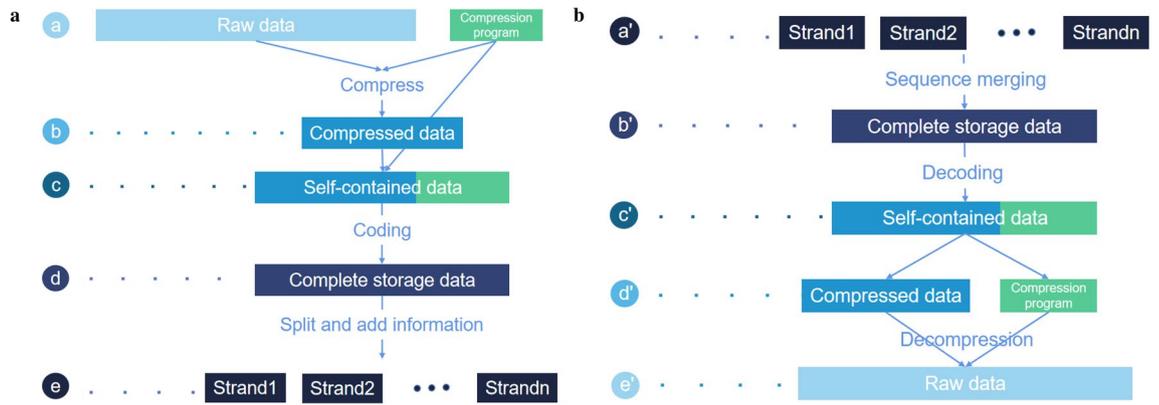

**Figure 2.** DNA storage process. (**a**) Shows how the dataflow is changed at each stage of the process. The compressed data (**b**) and the decompression program constitute the self-contained data (**c**), which can be simply combined into a single continuous file. However, this change might cause some problems, we will describe them in details later. After encrypting, encoding, and error correcting, we can obtain a long DNA sequence (**d**), which is then split into a set of short sequences (**e**), which are finally stored through the artificial synthetic process. In contrast, (**b**) shows how the dataflow change with respect to the reading process. The fragments are first merged into a complete long sequence (b'), which is then decoded to obtain the corresponding self-contained data (c'). The self-contained data is split into compressed data and a compression encoded program (d'), which is used to decompress the compressed data to obtain the original data (e')

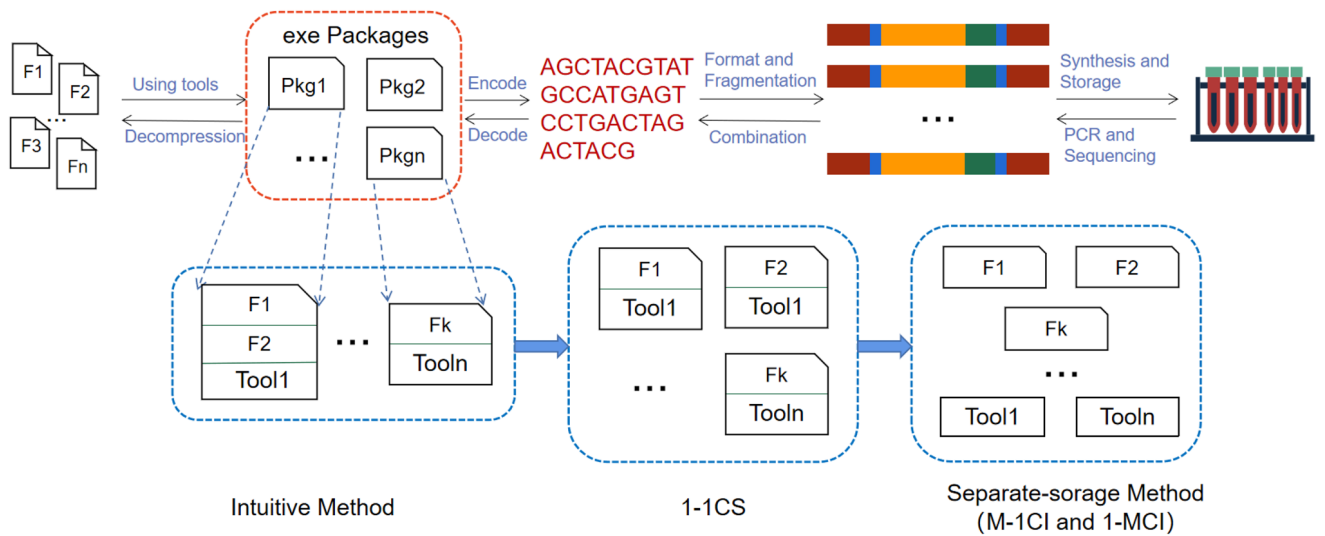

**Figure 3.** Storage process and three implementation methods: The general storage processing of DNA files is improved with the self-contained method. The red dashed box represents the design of file format while the first blue dashed box below represents *intuitive method*: combining the data files that adopt the same tool (e.g., using the same compression program) together. However, this method is not well suitable for DNA storage. The second blue dashed box indicates that each file is packaged separately which we call this method *one-to-one continuous storage* (1-1CS) method—the data and tools are stored together. In contrast, the third box represents discontinuous storage method where the tools and data are stored separately (separate-storage method), which is desired to minimise the data redundancy.

Therefore, we hope that the data and related tools stored in DNA are as complete as possible, allowing the data to be self-contained.

The purpose of our self-containment and self-explanation is to make the DNA file contain more information, not only including the encoded data itself, but also maintaining the index information for the stored data. As shown in Figs. 3, 4 and 5, we take the data compression in DNA storage as an example to explain the proposed self-contained and self-explanatory technology. In this example, the decompression program is our tool. As usual, the DNA fragments are stored in pools without specifying their orders, and all the stored files are identified using primers.

*Intuitive method and 1-1CS method.* The most intuitive method is the one shown in Fig. 3's first blue dotted box where as with the idea of executable compressed file, the different data files, together with their used program tools, are all packed into a single package. For instance, `F1` and `F2` are packaged into a self-extracting program file using compression tool `Tool1`, and similarly, `Fk` is packaged into a self-extracting file using compression tool `Tooln`. Although this method is easy to implement, it suffers from the drawback for DNA storage to lose the random readability—if one file (say, `F1`) is required, the entire package has to be read and decompressed to obtain the required file. This is difficult to work with DNA storage as the time and monetary cost of DNA sequencing could be much longer and expensive than those of data read from hard drives. As such, we have to enable the DNA storage system to support random read operations for specific compressed files. To address this issue, each file can be compressed independently as shown in the second blue frame where the decompression tool is appended to each compressed data, which we called *1-1CS* method, where "CS" stands for continuous storage. Although it can solve the random read issue, this method is still wasteful in space since if multiple files (say, `F1` and `F2`) use the same tool, and the tool has to be saved multiple times, one for each file. Therefore, to minimise the space cost, we need to reconfigure the executable compressed package by saving the data and the tool program in DNA *separately* as shown the last box in Fig. 3 and then set up the relationships between the data and its corresponding tool program for not only recovering the data with minimum sequencing cost but also supporting random read operations. We call this *separate-storage (SS) method*.

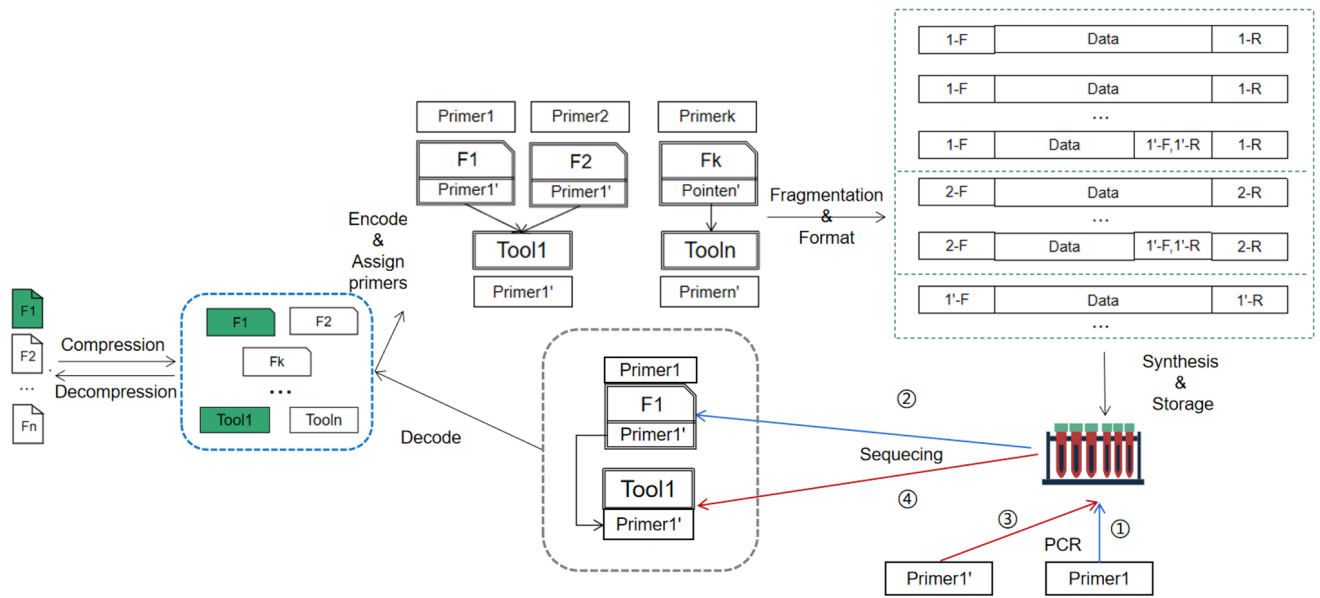

**Figure 4.** M-1CI Method: As with the previous method, data is compressed before it is stored. However, the compressed data files are not continuously stored in a single file as shown in the blue dashed box, instead, they are compressed into separate files. Next, for random read access, we index a file by adding a file primer to it (say, `Primer1` to `F1`), encode the file into a base DNA sequence, then append the tool's primer to the data file (say, `Primer1'` to `F1`), and finally, perform the sequence segmentation and add tool's primers to each fragment. Suppose in the reverse process if file `F1` is read, Step ① is to amplify it by `Primer1`, then in Step ②, `Primer1'` of `Tool1`, together with `F1`, can be obtained after a round of sequencing. Then, by using `Primer1'`, the tool file is amplified in Step ③, followed by Step ④ to obtain the `Tool1` file by another round of sequencing. After decoding, `Tool1` is used to decompress `F1` to obtain the original data. Overall, the sequencing procedure is performed twice.

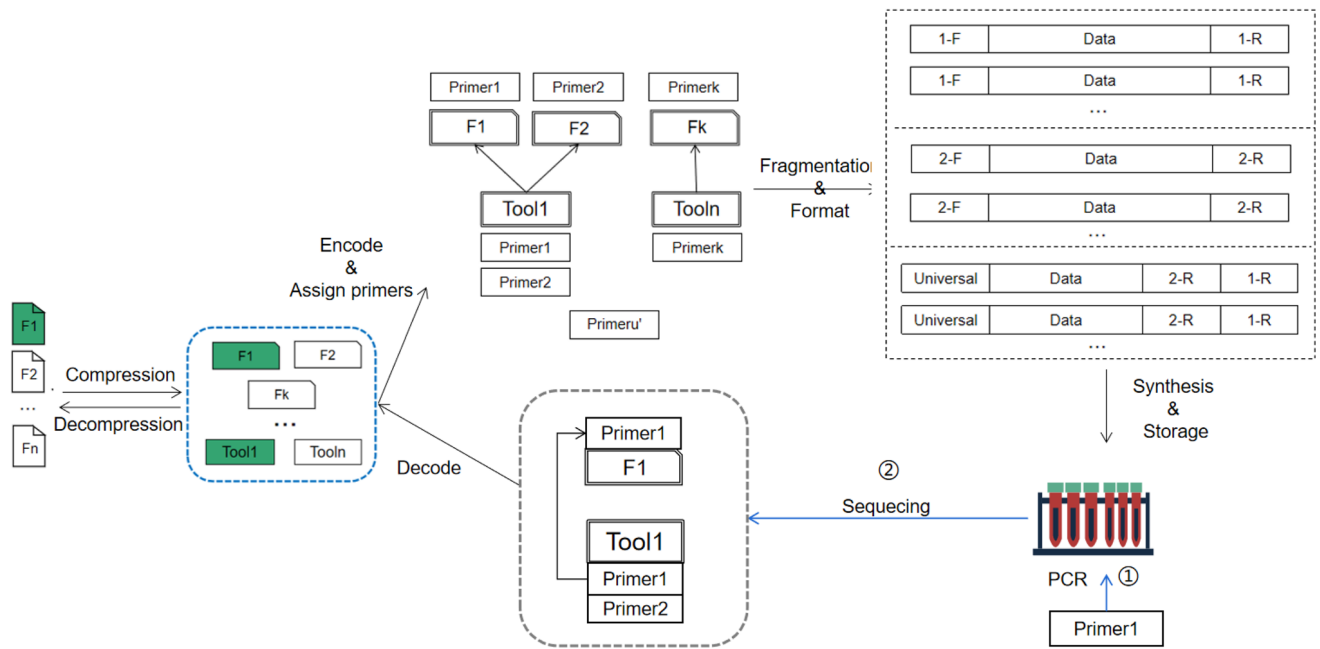

**Figure 5.** 1-MCI Method: The procedure prior to assigning primers is the same as that of M-1CI. Logically, 1-MCI no longer assigns a specific primer to each tool, rather, all the tools share the same universal primer and include all the primers used in their own data files. The specific implementation is reflected in the sequence after the segmentation is performed, the fragments of all the tool files contain the data file primers and the universal primers of the tools applied to the data files. Suppose if `F1` needs to be read, Step ① is to amplify it by `Primer1`, then in Step ②, sequencing, decoding, and decompressing can be performed once to restore the original file.

Basically, there are two basic ideas to design the SS method. One is to allow each of the data files being manipulated (say, compressed) with the same tool to have a pointer pointing to the same manipulation (say, decompression) tool file in DNA as shown in Fig. 4. The other is to enable a tool file to *reversely* point to multiple data files related to it as shown in Fig. 5. We call these two sub-methods *Many-to-One chain indexing* (M-1CI) method and *One-to-Many chain indexing* (1-MCI) method, respectively, which are described and compared in details in the sequel. In particular, we also show how 1-MCI is better than M-1CI in terms of sequencing cost.

*M-1CI method.* As described above, the basic idea of the M-1CI method is to allow each of the data files being manipulated with the same tool to have a pointer pointing to the same manipulation tool file in DNA with an attempt to minimise the space cost. More specifically, given the current used synthetic DNA for data storage is generally below 200 nucleotides, the stored file has to be segmented into a set of short fragments[17,19,20,22], each being marked with a pair of primers at both ends to indicate the data stored in between. In the design, the data file holds a pointer (i.e., the tool's primers) pointing to the tool file that can be used to resolve the fragment data. The pointer can be obtained by adding the primers of the tool file to the end of the data file as shown in Fig. 4. As such, the method is fairly effective in minimisation of the space cost since the separately stored tool used to manipulate a set of files can be found and retrieved only once whenever one of those file is randomly read. However, some practical issues make it difficult to realise the method in an effective way given the PCR (Polymerase Chain Reaction) and sequencing process in the DNA storage. In particular, this design may lead to so-called *double sequencing* problem—the synthesised DNA file data needs to be first sequenced to obtain its manipulation tool file's information, and then, by which the corresponding synthesised tool file data can be derived via sequencing to resolve the original data file.

To fully understand the problem, we use an example as shown in Fig. 4 where we assign `Primer1` to `F1`, `Primer2` to `F2` and `Primer1'` to `Tool1`. Here, `PrimerX` represents a pair of primers X-R and X-F at both ends of the sequence, for example, `Primer1` represents a pair of primers 1-F and 1-R assigned to `F1`. After the file is encoded into a long DNA sequence, it is then divided into small fragments. At this stage, each pair of primers is actually added to both ends of the file fragment. The big dashed box in green shows the design of the primer formats of `F1`, `F2`, and the tool being used. Each small box represents a collection of file fragments, where the `Data` field represents the file data (the detailed format will be introduced later).

For example, the first box represents the fragment belonging to `F1`. The pointers `1'-F` and `1'-R` of `F1` to file `Tool1` are included in the data field of `F1`, and may be at an end of a fragment. If the data of `F1` needs to be read, we need to perform the PCR amplification for the sequence of `F1`, according to the known primer `Primer1`. In the second step, the data of `F1` is obtained, and then `Primer1'` of `Tool1` is obtained. In the third step, the previous steps are repeated to amplify and sequence to get the `Tool1`'s data according to `Primer1'`, and finally use `Tool1` to process `F1` to restore the original data (e.g., decompress). With this method, one has to restore the data file with 2-round sequencing operations, so-called double sequencing, in which the first round sequencing is to use the selected primers to amplify the specified data file and then sequence it to obtain the primers of the tool file. As the DNA sequencing is time-consuming and costly, this double sequencing method is not effective in both time and expenditure.

*1-MCI method.* In response to the *double sequencing* problem, we propose a multi-primer-based method, named *One-to-Many Chain Indexing* (1-MCI) method, which uses separate storage for data file and tool file, and tweaks their respective primers as the pointers to each other with an aim to have only one round sequencing to restore the complete data file as shown in Fig. 5. To this end, we deliberately modify the data format of the DNA fragment so that the tool file can have a reverse pointer pointing to its data file. In contrast to M-1CI, the 1-MCI method has the distinct feature that allows the tool file to have all the primers of the data files that have been manipulated by the tool, instead of its own primer as does in M-1CI. With this design, 1-MCI can restore the selective data file with only one round sequencing as the data file and its manipulation tool file are available at the same time, which, compared with M-1CI, dramatically reduce both time and expenditure in data restoration. Note that in this procedure, we also need to distinguish the data file and the tool file by assigning a universal primer to the tool file.

An illustrative example is shown in Fig. 5 where we first assign `Primer1` to `F1`, `Primer2` to `F2`, and `PrimerK` to `Fk`, and then allow `Tool1` file fragments to include both `Primer1` and `Primer2`, `ToolN` to include `PrimerK`. Finally, we enable all the tools to share a common known primer `PrimerU'` as the universal primer. The detailed design of the primers after `F1`, `F2` and `Tool1` segmentation is shown in the large dashed box in Fig. 5. Now, the data of `F1` and `F2` no longer contain the primer of `Tool1` as in the 1-MCI method, instead, all the fragments of `Tool1` include sub-primers of `Primer1` (`1-F` or `1-R`) and `Primer2` (`2-F` or `2-R`) and a universal primer `PrimerU'`, which are used to selectively amplify the file to be read as well as its associate tool. Specifically, with this design, when `F1` is selectively read at this time, we only need to amplify the sequence based on the pre-known `F1`'s primers `Primer1` (`1-F` and `1-R`) and preset the universal primers, and then get `1-F` and all the data fragments identified by `1-R` belonging to `F1` and all the fragments of `Tool1` identified by `PrimerU'` and `1-R`, thereby directly obtaining the original data with only one round sequencing.

| Field name | Address offset | Size | Description |
|---|---|---|---|
| FT | 0 | 1B | File type |
| FID | 1 | 4B | File identifier |
| SM | 5 | 1B | Storage method |
| DL or TFID | 6 | 4B | Data length or Tool file's FID |
| D | 10 | Variable length | Length of data |
| TD | Variable length | Variable length | Tool's data |

**Table 1.** Data file format. `FT` means file type, which is either data file or tool file. `FID` uniquely identifies a file with an 1-to-1 mapping to its file name. `SM` represents three kinds of storage methods—*unprocessed file* (`OF`), *continuously processed file* (`CPF`), and *non-continuously processed file* (`SPF`). If a file is `OF`, field `D` stores its data, otherwise, if it is `CPF`, then the next field is `DL`, specifying the length of the processed data, if it is `SPF`, the next field is `TFID`, which means the `FID` of the used tool file. The last field `TD` is variable length, if it is `CPF`, `TD` means tool data, else the last field length is set to 0.

| Field name | Address offset | Size | Description |
|---|---|---|---|
| FT | 0 | 1B | File type |
| FID | 1 | 4B | File identifier |
| TD | 5 | Variable length | Tool's data |

**Table 2.** Tool file format. As with the data file, the first field `FT` specifies file type, the second field `FID` is the identifier of the file, and the third field `TD` is a variable-length field, which stores the data of the tool.

The problem with this method is that artificially synthesised DNA fragments have a certain length limitation. As such, if the DNA fragment contains too many program file's primers, the effective data load for the tool file will be reduced. Much worse, if a large number of data files are manipulated using the same tool program, the DNA fragments of the tool files may not contain the primers of all the data files. Given this consideration, it is necessary to strike a balance between the self-contained data and the data redundant overload. Overall, even with this problem, the proposed 1-MCI method is still of great significance in practice not only for the correspondence between the tool programs and the data files, but also for other types of file indexes in the DNA storage.

**Data self-explanation.** The goal of the data self-containment is able to restore the compressed data stored in the DNA storage without relying on external decompression tools. In particular, when reading a data file, the system first finds the file and the primer sequence corresponding to its decompression tools, and then obtains the data and the tool files at the same time through the PCR and DNA sequencing technology. After decoding, the decompression tool can automatically restore the data file to its original form to realise the self-explanatory function of the data. Clearly, to achieve this goal, one has to embed many different kinds of information into the DNA file in such a way that the self-contained data is also sufficient to self-explanation, which requires a well-defined file format.

*Data file format for self-explanation.* The definition of the data format is to support multiple implementations of data self-containment with the ability of self-interpretation. The data format should include two levels of metadata information. The first level of the metadata describes a format for the binary compressed file in the data pre-processing step while the second level of the metadata defines the format for those small fragments after the file is encoded into a DNA sequence.

**DNA File Format:** The last steps in Figs. 4 and 5 are performed to select the specified data files and and their associated tool files. If there are multiple files selected to read at same time, a certain data format is required to support the realisation of data self-extraction for these files without compromising others. To this end, we define the format of data file as shown in Table 1.

In our design, tool file and data file are stored in separate storage. The format of tool file is shown in Table 2.

We realise the data self-explanation mainly through the definition of the file format. The logic of file writing is relatively simple, which is simply to write the file in sequence, according to the field orders defined by the file format. The reading process is performed after the DNA data is decoded into binary data. First, the data file and

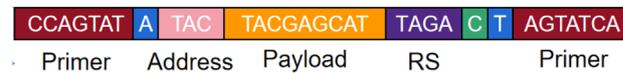

**Figure 6.** DNA fragment format. The head and tail are primer sequences used to amplify specific file sequences with the PCR technology. The `A` and `T` at the ends indicate the direction of the sequence, and `Address`[17] gives the offset of the data in the file. The middle `Payload` field is the data payload, and the `RS` field in purple is the error correction code. Since PCR can amplify the DNA fragments of the data file and the tool file at the same time, the sequencing result may contain both two files, so field `C` in green indicates whether the fragment belongs to the data file or the tool file.

---

the tool file are separated according to the first field `FT`, and the data file is put in the *datafiles* array, the tool file in the *toolfiles* array. Then, Algorithm 1 for file reading is performed, which first finds the needed file according to `FID` in the *datafiles* array (`Line 4-5`), then determines the storage method of the file according to `SM`. If `SM` is `OF`, the next field is the data field, and the subsequent field is data `D`, which can be read directly and returned (`Line 6-7`). Otherwise if `SM` is `CPF`, the data field `D` could be obtained based on the data length `DL` of the next field, and the subsequent data is the tool field `TD`, which can be used to restore the processed data to the original data and return (`Line 8-12`). Otherwise if `SM` is `SF`, the `FID` of the tool file can be obtained according to field `TFID`, then the tool files array can be traversed to find `TD`, whereby the processed data can be restored to the original data (`Line 14-20`).

---

**Algorithm 1** *ReadFile*

**Input:** $FID, datafiles, toolfiles$
**Output:** $filedata$
 1: $filedata \leftarrow None$
 2: $datalength \leftarrow 0$
 3: $toolfilename \leftarrow None$
 4: **for** (each $a \in datafiles$) **do**
 5:   **if** ($a.FID = FID$) **then**
 6:     **if** ($a.SM = OF$) **then**
 7:       $filedata \leftarrow a.data$
 8:     **else if** ($a.SM = CPF$) **then**
 9:       $datalength \leftarrow a.DL$
10:       get the processed D field through *datalength*
11:       get the TD field in the remaining data
12:       use TD to recover D and get *filedata*
13:     **else**
14:       $TFID \leftarrow a.TFID$
15:       **for** (each $t \in toolfiles$) **do**
16:         **if** ($t.FID = TFID$) **then**
17:           get the TD
18:           use TD to recover D and get *filedata*
19:         **end if**
20:       **end for**
21:     **end if**
22:   **end if**
23: **end for**
24: **return** *filedata*

---

**DNA Fragment Format:** Due to the limitation of the length of the artificially synthesised DNA sequence, it is necessary to divide the encoded long DNA sequence into small fragments for synthesis. This small DNA fragment consists of multiple fields, as shown in Fig. 6.

**DNA storage system architecture.** The overall architecture of the DNA storage system is illustrated in Fig. 7. The data files stored on traditional storage medium is firstly pre-processed. The pre-processing procedure is composed of data compression, data deduplication, and data formatting, which follows the defined DNA segment format for self-containment. The proposed algorithm then encodes the data into DNA sequence for DNA synthesis. The synthesised DNA is stored in DNA storage medium and sequenced to produce the DNA sequence. The DNA sequence is then decoded into binary files. Finally, the compression algorithm that is self-contained in the sequence is used to extract the file content, achieving the self-explanatory.

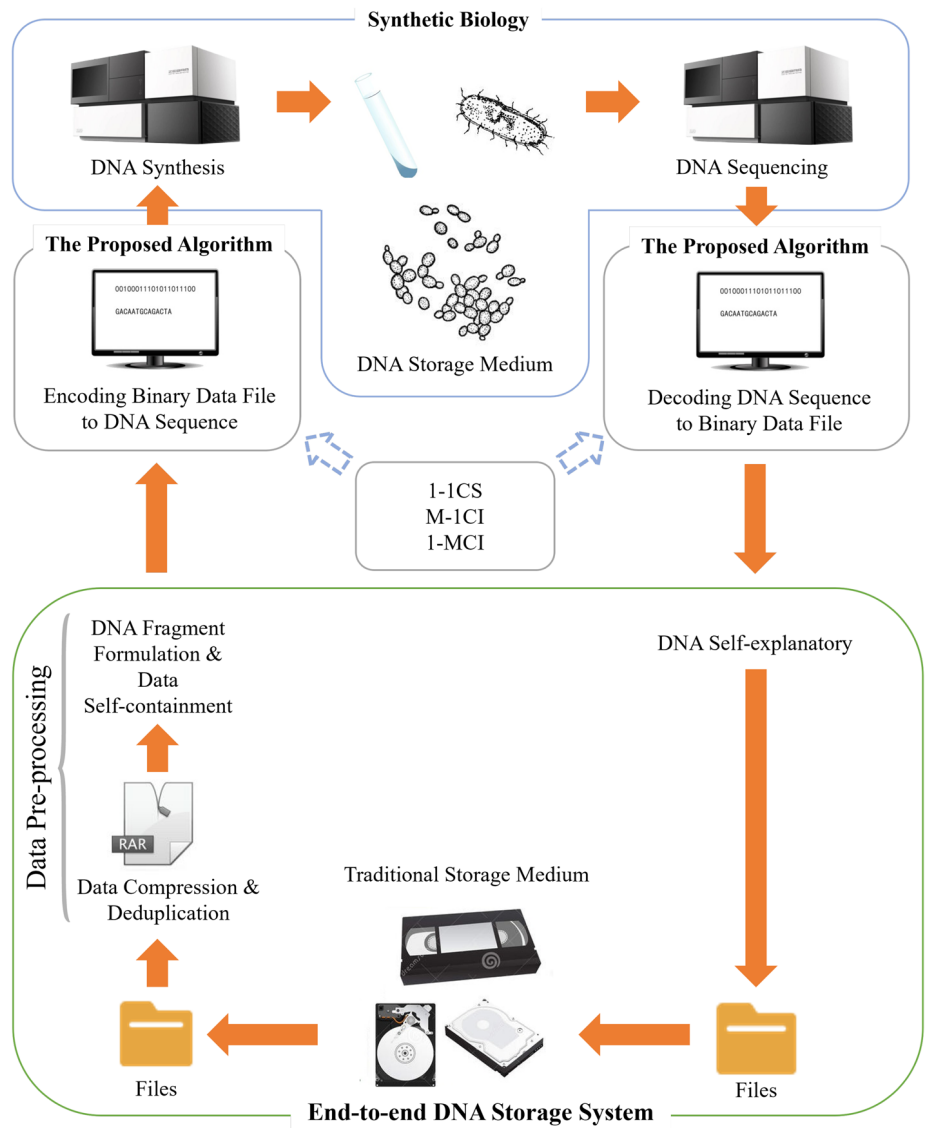

**Figure 7.** Overall DNA storage system architecture: The data files stored on traditional storage medium are pre-processed and handled by the proposed algorithm before being synthesised. The synthesised DNA can be restored by being sequenced to produce the original binary data.

### Results

As shown in Fig. 2, the DNA storage process can be in general divided into two inter-connected parts—a digital part and a bio-laboratory part, and this research is conducted with a focus squarely on the digital part. One of the key problems in this part is how to integrate external tools into the encoded DNA sequence for high storage density. To evaluate the effectiveness of our approach, we select three often-used compression programs—*rar*, *7z* and *zip*–as the tool and a set of different types of data as the target data to be stored. The used symbols are list in Table 3 in "Appendix".

**Storage overhead.** The data self-containment inevitably brings some storage overheads. To evaluate this impact, we first define *compression efficiency*, denoted by *e*, as the metric to measure the space impact of the selected compression programs in the worst case, whereby the program with the best performance is selected as the main test program to evaluate our proposed methods.

Formally, the compression efficiency is defined as $e_c = 1 - e$ where $e$ is given by $e = (r_c \times S_o + S_T)/S_o$, here, $S_o$ represents the original data size, $S_T$ is the size of the tool file, and $r_c$ is *compression ratio*, defined on per-program basis as $r_c = S_c/S_o$ where $S_c$ is the size of the file after it is compressed. Given this definition, we can make two key observations.

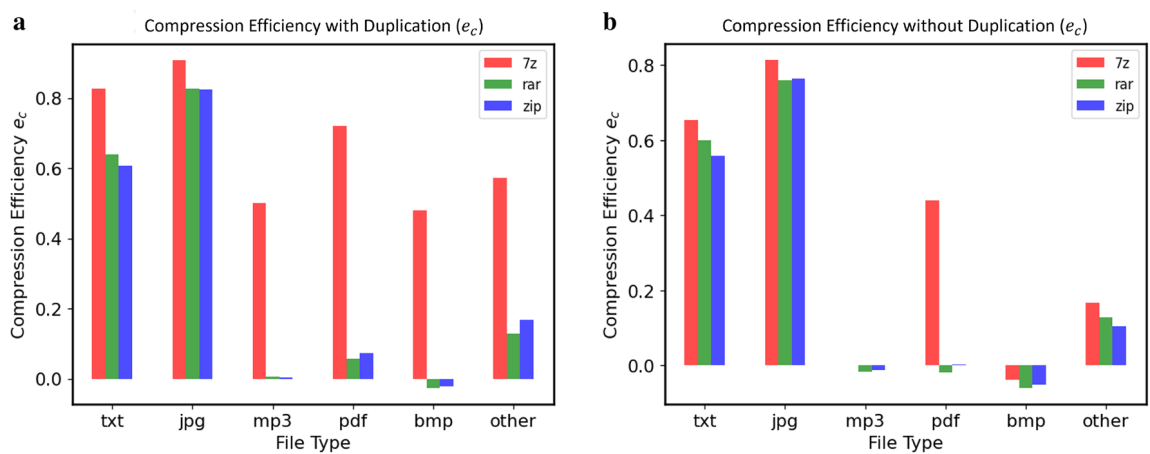

**Figure 8.** Compression efficiency (data types). This figure shows how the $e_c$ values of the three selected compression methods are changed with respect to six types of test data (*txt*, *mp3*, *jpg*, *pdf* and other files, such as *exe*, *dll*, *html*, etc.) for the cases—the data is either redundant (**a**) or not (**b**). The data sizes used by these two figures are 4*MB* (w/ redundancy) and 2*MB* (w/o redundancy), respectively. In particular, (**a**) shows that the overall compression efficiency of *7z* is best among the three, and the relative effects of *rar* and *zip* are different in different situations. From (**b**), one can still see that *7z* is the best, but its advantage is diminished when the data redundancy is absent.

**Observation 1** Not only compressed data but also external tool are considered in the computation of $e_c$. In particular, when no self-containment is involved, i.e., $S_T = 0$, $e_c$ is reduced to $e_c = 1 - r_c$, which represents an ideal case for overall compression effects.

**Observation 2** It is not always beneficial to use the self-containment method for the DNA storage as $e_c$ could be less than or equal to 0 when $S_T \geq (1 - r_c)S_o$. In other words, $e_c$ can provide a guidance in the use of our method in the worst case.

The compression efficiencies of the selected programs with respect to different types of data files are compared in Fig. 8 where compression efficiency $e_c$ vs. 6 common types of test data files are depicted. In the figure, the storage overhead $S_T$ of *7z* is about 201 KB, *rar* is 303.5 KB and *zip* is 254 KB, and the test data with and without redundancy are 4*MB* and 2 MB, respectively. One can see that the higher the value of $e_c$, the better the compression effects, which means that the space overhead of compression program is smaller and smaller. On the other hand, it is also validated that the self-containment method in the worst case is not always beneficial as shown in the figure for the *bmp* file.

Self-contained data will inevitably bring in space overhead (i.e., $S_T > 0$). Given the results of Fig. 8, we deliberately choose *7z* as the compression tool and *txt* as the test data due to their relatively good performance, and then conduct a test to measure the overhead of the proposed self-contained and self-explanatory method. The results of the test are shown in Fig. 9 in which how the values of $e_c$ in *ideal case* (in blue) and in self-containment case (in red) are changed with respect to $S_o$ and $r_c$ are depicted.

It can be seen from Fig. 9 that the self-containment method is not always beneficial to saving storage space. Rather, it is possible to incur extra storage overhead, compared to the default method that no compression is employed. However, when the amount of data reaches a certain threshold, the self-containment method approaches those methods in ideal case, and the storage overhead incurred by the self-containment at this time can be almost ignored.

**Data storage.** In contrast to the previous section, which focuses on the compression efficiencies of selected compression programs in the worst intuitive and 1-CS cases, we evaluated the proposed 1-1CS, M-1CI and 1-MCI in this section when they are used to achieve self-containment and self-explanation. For fair comparison, we deliberately assume that all three methods are used to store an equal number of $n$ data files ($D_i, i \in 1, 2 \ldots n$). Due to the limitation on the length of synthetic DNA fragment, the range of $n$ is thus determined by 1-MCI and limited to $0 < n < \lfloor L_s/L_p \rfloor$ where $L_s$ represents the maximal length of DNA segment in nts and $L_p$ the length of primer in nts. We first analyse the synthesised and sequenced amount of data with respect to each proposed method in theory, and then carry out a DNA data storage process on a real platform to validate the conclusions made above.

*Method analysis.* With the analysis in mind, we first give the amount of binary data $Sb_M$ that needs to be stored by 1-1CS, 1-MCI and M-1CI as follows:

$$Sb_{1-1CS} = \sum_{i=1}^{n} r_c \times S_{D_i} + n \times (S_h + S_T) \quad (1)$$

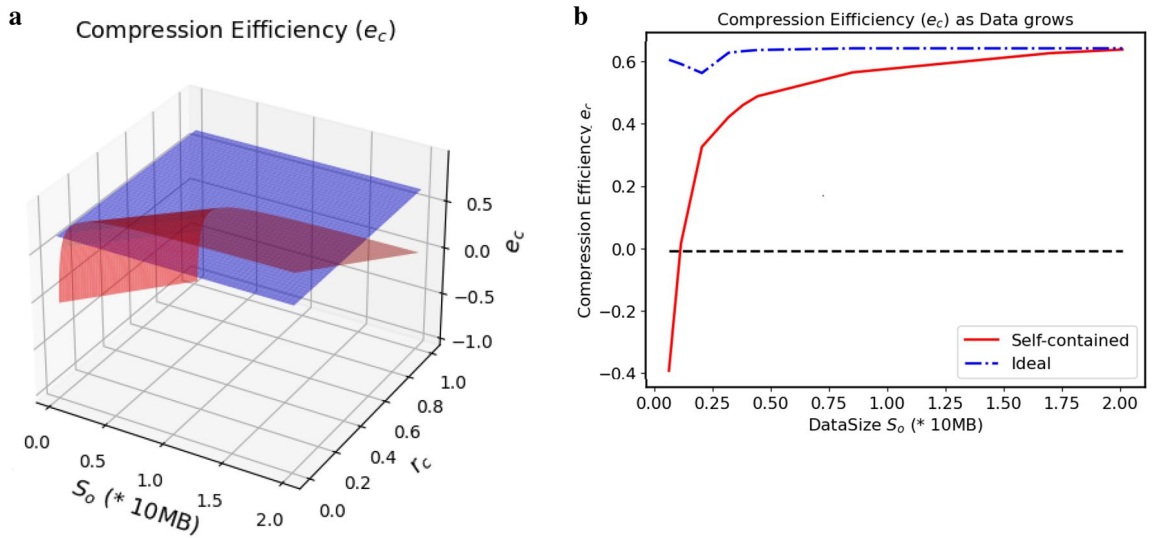

**Figure 9.** Compression efficiency (data size). We calculate the compression efficiency of *7z* when the size of the original data $S_o$ gradually increases. Here, $r_c$ of *7z* is 0.34 when a text file in size of 500 MB is compressed. Given that $S_T$ of *7z* is 201 KB, we have the results as shown in (**a**) where $e_c$s are calculated for the data storage method in *ideal case* by $e_c = 1 - r_c = 0.66$ and in the self-containment case by $e_c = 1 - (r_c \times S_o + 201K)/S_o$. To validate the results, we further pack a set of test data files with different types (e.g., *txt, jpg, exe, mp3*) in a single compressed package and compare their actual values of $e_c$ with a fixed $r_c = 0.34$ as shown in (**b**). The results are roughly consistent with those in calculation as shown in (**a**).

$$Sb_{M-1CI} = \sum_{i=1}^{n} r_c \times S_{D_i} + n \times (S_h + 2L_p) + S_T \quad (2)$$

$$Sb_{1-MCI} = \sum_{i=1}^{n} r_c \times S_{D_i} + n \times S_h + S_T. \quad (3)$$

In addition to the 1-MCI tool files, the DNA fragments of other files contain a primer at both ends, so the length of the data of the fragment is $L_s - 2L_p$, the DNA fragment of the 1-MCI method' tool file contains primers for $n$ data files and a universal primer, so the length is $L_s - (n+1)L_p$. The total number of bases $Sd_M$ after adding the primers to the segmented fragments produced by three methods are:

$$Sd_{1-1CS} = Sb_{1-1CS} \times a + \frac{Sb_{1-1CS} \times a}{L_s - 2 \times L_p} \times 2L_p = a \times Sb_{1-1CS} \times (1 + \frac{2L_p}{L_s - 2L_p}) \quad (4)$$

$$Sd_{M-1CI} = Sb_{M-1CI} \times a + \frac{Sb_{M-1CI} \times a}{L_s - 2 \times L_p} \times 2L_p = a \times Sb_{M-1CI} \times (1 + \frac{2L_p}{L_s - 2L_p}) \quad (5)$$

$$Sd_{1-MCI} = Sb_{1-MCI} \times a + \frac{(\sum_{i=1}^{n} r_c \times S_{D_i} + n \times S_h) \times a}{L_s} \times 2L_P + \frac{S_T \times a}{L_s - (n+1)L_p} \times (n+1)L_p \quad (6)$$

where $a$ is the base factor.

In order to simplify the calculation, we assume that each $D_i$ has an equal size $S_D$. Figure 10 presents the quantitative results of the number of bases that need to be stored for each method, i.e., $Sb_M$, under different settings. In Fig. 10a, we can observe that all three methods show a clear linear trend, and the 1-1CS method requires the highest DNA base storage capacity compared with its counterparts. Both 1-MCI and M-1CI show a relatively lower DNA base storage capacity, which demonstrates the effectiveness of using primers to avoid data redundancy. Figure 10b presents the trend of these three methods when the $L_s/L_p$ ratio varies. A clear decreasing trend is observed, and the descending speed slows down when the ratio becomes larger. Again, the 1-1CS requires the highest amount of DNA storage, followed by the M-1CI. 1-MCI achieves the lowest DNA storage due to the effectiveness brought by the data redundancy avoidance.

Generally speaking, when these three methods are being compared, the 1-1CS has the largest storage overhead since it needs to store multiple copies of the compressing tool. By avoiding this data redundancy, the 1-MCI and M-1CI perform better than 1-1CS in terms of $Sd$.

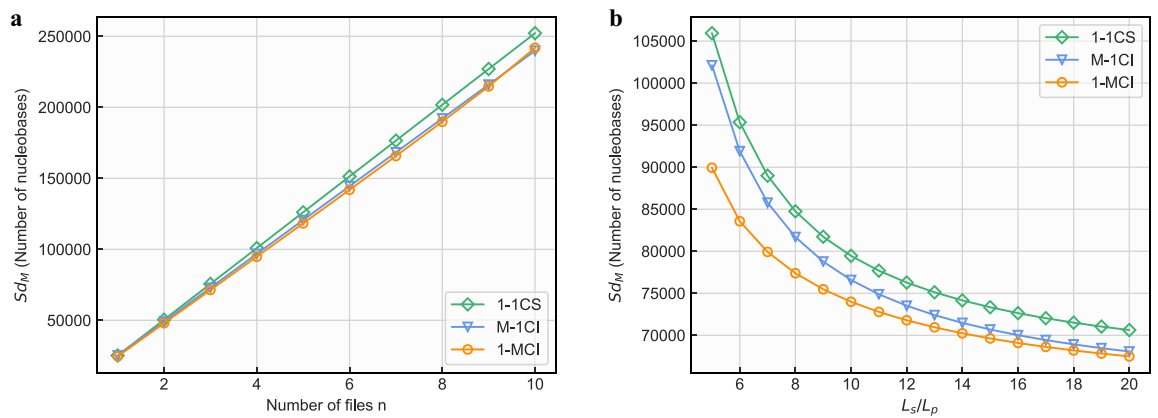

**Figure 10.** Number of bases that need to be stored for each self-containment method. (**a**) Illustrates how the $Sd_M$ changes when the number of files n varies. All three methods show a clear linear trend, and the 1-1CS method requires the highest DNA storage capacity compared with its counterparts. Both 1-MCI and M-1CI show a relatively lower DNA base storage capacity. (**b**) Presents the trend of $Sd_M$ when the $L_s/L_p$ ratio changes. A clear decreasing trend is observed, and the descending speed slows down when the ratio becomes larger. Again, the 1-1CS requires the highest amount of DNA storage, followed by the M-1CI. 1-MCI achieves the lowest DNA base storage due to the effectiveness brought by the data redundancy avoidance. Note that we set the parameters $S_h = 10, S_T = 2010, L_p = 20$, and $a = 1/1.6$ during experiments. $S_D$ is set to be 100000 to make sure that the size of data far exceeds the size of primers. $r_c$ is 0.34 for the 7z compression algorithm. $L_s$ is also greater than $L_p$, it varies between 100 and 400 to achieve a ratio ranging between 5 and 20.

*Method tests.* As opposed to the above studies which focus on the theoretical calculations, we carried out a DNA data storage process on a real platform to validate the conclusions made above. The setup of our experiment is configured with a Intel(R) Core(TM)-7-9700 CPU@3.00GHz processor, the RAM memory with size 16.0 GB, the OS configuration is 64-bit Windows10, and the experimental environment is Python3.8. Since our method is mainly to design the indexing method of files and related tools, the design of the coding algorithm is not what we are concerned about. We used the open source tool Chamaeleo[28] to test our method.

In this experiment, we selected two compression program tools, *7z* (194KB in size) and *zip* (87KB in size). Although *7z* is relatively large, its compression effects for those files with more redundant data is better than those of *zip*. We established 5 file directories—Cat, Jane, MonaLisa, Leo, and mix to store the data files in 3 types—bmp, txt, and jpg—as independent data items for testing. The references of the dataset we used are given in the "Appendix" section.

**Experiment design:** As discussed, in this paper we are mainly concerned with the digital part of the DNA storage process, the actual bio-synthetic storage is deliberately ignored. Similarly, we assume that the data can be successfully amplified by PCR and sequenced, and then decoded and decompressed after it is read out to realise its self-interpretation.

*(a) Primer Design:* Since the file indexing in our method mainly relies on the allocation of primers, we also carried out different primer designs for the experiments. When designing the primers, we should not only consider the parameters of biological information that need to be followed, but also concern with the issue of homology[15]. Considering the impact on time and space, we selected *Grass*[17] encoding DNA files for the primer design. We thus needed to design a pair of primers for each data file. The data files of all the methods need a pair of primers. For 1-1CS and M-1CI, we needed to design a pair of primers for each tool file. For 1-MCI, we only needed to design a pair of universal primers (PrimerU') for the two tool files. Finally, adding primers to the data fragment realises the calculation phase of data self-containment.

For both 1-1CS and M-1CI, there are primer sequences on the left and right ends of the fragments in both data file and tool file. We set the primer sequence with a length of $L_P = 20$, the effective data load thus will not exceed $220 - 40 = 180$. Taking into account the space required for the index, we set the final data load to be $160bp$, and the space occupied by the index varies according to the file size. Similarly, for 1-MCI, the two ends of the DNA fragment of the data file are still primer sequences, and the data payload is also $160pb$. However, for the DNA fragment of the tool file, it needs to include the primer of the data file and a universal primer, so its *7z* and *zip* data payloads are thus up to $220 - 2 \times 20 - 20 = 160$ and $220 - 3 \times 20 - 20 = 140$, respectively. Considering the space occupied by the index, we set the final data length for *7z* and *zip* to 140 and 10, respectively. The index length changes based on the file size.

*(b) Process design:* We first use different compression programs to compress a set of test data (5 data files in different types and 2 compression programs), and then adopt a classic coding scheme to encode the compressed data based on the Chamaeleo software. More specifically, we assume that the length of the synthetic DNA sequence is within $220bp$, and then make the encoding process that is first to read in each test file as binary data, then divide it into fragments, add an index to identify the position of fragment, and then exploit the selected encoding algorithm to encode it into a DNA sequence, and finally add the designed primers at its both ends. As the error problem is not the focus of this paper, we do not add an error correction code field to the fragment.

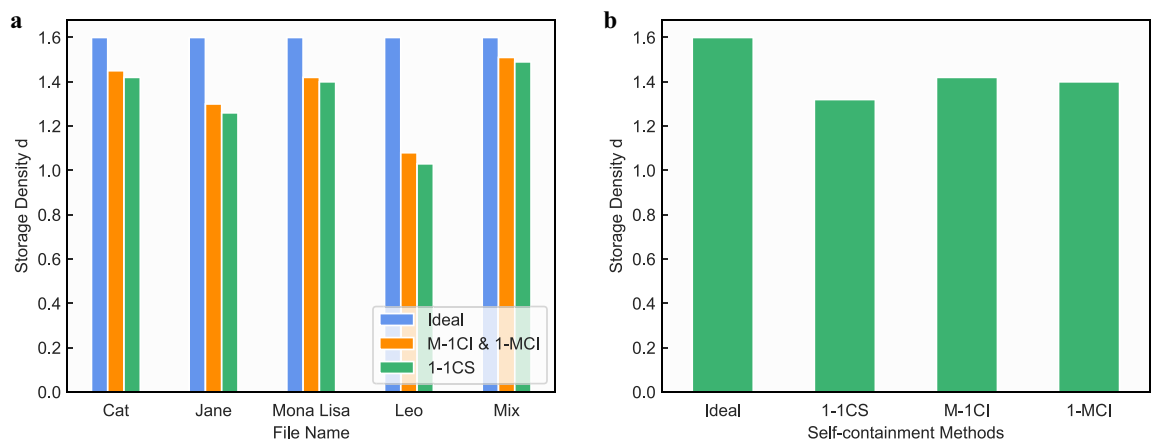

**Figure 11.** Storage density. (**a**) presents the storage density $d = S_o/S_b$ when storing different files using three methods. In the ideal case, only the compressed data will be stored and its compressor will be ignored. Note that the M-1CI and 1-MCI methods present the same storage density performance, hence they are represented using only one bar. Thanks to the avoidance of the data redundancy, the 1-MCI and M-1CI method demonstrate higher storage density compared with the 1-1CS. (**b**) shows the storage density of these methods when storing all the data. Again in this case, both 1-MCI and M-1CI present nearly the same storage density performance, which outperform the 1-1CS method.

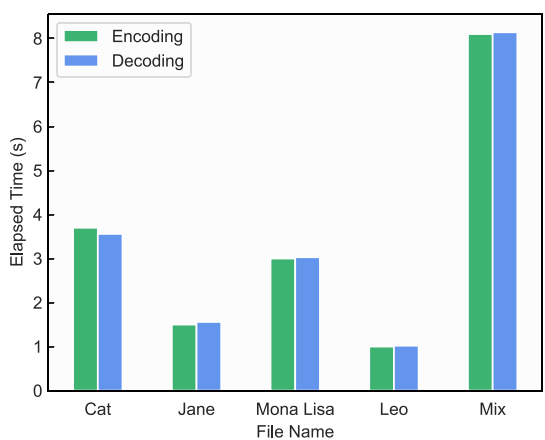

**Figure 12.** Time efficiency. This figure presents the time taken for the encoding and decoding process when processing different files under 1-MCI and M-1CI storage methods. Since 1-MCI and M-1CI present nearly the same time performance, therefore, the bar chart does not present different methods as one extra dimension.

**Experimental results:** With the above designs, we store the five test files in the DNA storage with different proposed methods and then measure their respective storage densities and time taken to reveal the advantages of the proposed methods. The storage density is defined as $d = S_o/S_b = 1/a$, here, $S_b$ is the total number of bases obtained by the selected encoding scheme, $a$ is the base factor. This metric indicates how many bits, including all the data, tool, and their respective index primers, are represented by a DNA base, which reflects how effective the self-containment methods can be used with DNA coding schemes. According to this definition, the higher the storage density, the better the method. The relevant results are shown in Fig. 11 where the storage densities of five different types of test files are compared, these files are first compressed by *7z* and then encoded with the coding algorithm in Grass et al.[17], and finally stored with the proposed self-contained and self-explanatory methods. In Fig. 11a, the storage density $d$ when storing different files using these three methods are presented. In the ideal case, only the compressed data will be stored, and it compressor tool will be ignored. By avoiding the data redundancy, the 1-MCI and M-1CI method demonstrate higher storage density, compared with the 1-1CS method. Hence, it indicates the effectiveness of using primers to reduce redundant data storage and boost the storage density. The same result is also supported by Fig. 11b, which shows the storage density $d$ of these methods when storing all the data files. Again, both 1-MCI and M-1CI outperform the 1-1CS in terms of storage density. In Fig. 12a, it presents the time taken for the encoding and decoding process when processing different files under 1-MCI and M-1CI storage methods during the digital part. As we can observe, both the 1-MCI and M-1CI present nearly the same reasonable time performance, which verifies that the proposed method runs reasonably when processing data files.

| Symbol | Description |
|---|---|
| $r_c$ | Compression ratio |
| $S_o$ | Size of original data before being compressed (B) |
| $S_c$ | Size of compressed data (B) |
| $e$ | Compression efficiency |
| $e_o$ | Compression efficiency of non-self-contained data |
| $e_c$ | Compression efficiency of self-contained data |
| $L_p$ | Length of primer (bp) |
| $L_s$ | Max length of DNA segment (bp) |
| $D_i$ | Data file with label $i$, $i = <1, 2, \ldots, n>$ |
| $S_{D_i}$ | Size of $Di$ (B) |
| $S_D$ | Binary file size when all $D_i$s are with the same size (B) |
| $n$ | Number of data files |
| $S_h$ | Size of data file's header (B) |
| $T$ | Tool file |
| $S_T$ | Size of tool file (B) |
| $M$ | Data self-contained method, M = < 1-1CS,M-1CI,1-MCI> (B) |
| $Sb_M$ | Size of binary data needs to be stored of method M (B) |
| $a$ | Base factor (bp/bit) |
| $Sd_M$ | Number of base needs to be stored of method M (bp) |
| $S_e$ | The number of bases finally stored |
| $d$ | Data storage density |

**Table 3.** Symbol description: Explanation of symbols to be used

**Summary:** In summary, the self-contained and self-explanatory technology will bring certain data overhead, but it greatly improves the integrity of data, and also ensures the reliable storage of data in external unreliable environment. In general, when the tool files are relatively small, The 1-1CS method can be used. While in case of relatively large tools, the M-1CI method can be used to minimise the redundant data overhead and reduce the number of sequencing passes for cost reduction. However, if there are a number of data files using the same tool, one can adopt the 1-MCI method to achieve an effective solution.

## Discussion

The current DNA storage often resorts to external tool to restore the original data from the stored DNA sequence, compromising the data integrity, which is critical to ultra-long-term data archives. To address this issue, we proposed a self-contained and self-explanatory DNA storage in this paper, which not only contains the data to be stored but also includes the tool that is used in pre-processing of the data, say compression program. The proposed system needs to be well-designed to address two problems—data redundancy and random read. To this end, we developed the 1-MCI method to minimise the redundant data and defined file formats for both DNA and its fragment files to realise the self-explanation function. We evaluated the proposed system via a prototype implementation and analytical experiments, which showed that the redundancy brought by the self-contained data can be successfully avoided by the proposed method. Additionally, we also made an in-depth exploration of the file index design with a corresponding file and DNA fragment storage formats.

Although it is proposed to address the data integrity issue for compressed data restoration as a holy grail, our method is generic enough to address other issues that may also benefit the DNA storage. For example, embedding the metadata for the data deduplication[29] in the DNA storage could dramatically improve the system robustness in uncertain settings while increasing the cost-efficiency, which lay the foundation for the application of the self-contained DNA storage in large-scale data storage.

## Appendix

Data Sources:

cat.jpg: Original picture
7z: https://sourceforge.net/projects/sevenzip/
Zipfile.py: https://github.com/python/cpython/blob/3.9/Lib/zipfile.py
MonaLisa_bmp.jpg: https://en.wikipedia.org/wiki/Mona_Lisa.wiki
Jane_eye.txt: https://www.gutenberg.org/ebooks/1260
Resurrection_Leo_Tolstoy.txt: https://en.wikipedia.org/wiki/Resurrection_(Tolstoy_novel)

Generated primers:

Primer of Cat for 3 methods: CGCGTATATGGCCCCTCTTC, ACTAACCTTGCCGTCGCCCT
Primer of Jane for 3 methods: CCCCGGTCAAAAGACAACGT, AGGTATCTGCGGCCCATTAAC
Primer of Monalisa for 3 methods: CCAGTGTCCGACGAACTTATCT, AGGGTAGAGCTCGGCAATGT
Primer of Leo for 3 methods: CTTTGCAACCGATTTCCACGT, GCAGGAATCCGTGGCCTAAAG
Primer of Mix for 3 methods: ACCTCCAGACCCCAGCTAAT, TCCGACCTTCCCAGCTAAAC
Primer of 7z for 1-1CS and M-1CI: ATCGGGTCAAAGAGGCGAAG, GCAGGCAAGCTCGTCGACAT
Primer of zipfile for 1-1CS and M-1CI: GCCGGCCTTCACAACTACAG, ACGCATACCACCCGCATACT
Universal primer of 7z and zipfile for 3 1-MCI: GCACAGCATAGCGTCCCTTG, AGTGGGGTTGAGCGTCGAAC

## References

1. Reinsel, D., Gantz, J. & Rydning, J. Data age 2025: the digitization of the world from edge to core. *IDC White Paper Doc US44413318* 1–29 (2018).
2. Reinsel, D., Gantz, J. & Rydning, J. White paper: The digitization of the world from edge to core. Tech. Rep., Technical Report US44413318, International Data Corporation, Framingham (2018).
3. Bohannon, J. DNA: The ultimate hard drive. *Science* (2012).
4. Wiener, N. Machines smarter than men? interview with Dr. Norbert Wiener. noted scientist. *US News & World Report* 84–86 (1964).
5. Neiman, M. On the molecular memory systems and the directed mutations. *Radiotekhnika* **6**, 1–8 (1965).
6. Clelland, C. T., Risca, V. & Bancroft, C. Hiding messages in DNA microdots. *Nature* **399**, 533–534 (1999).
7. Bancroft, C., Bowler, T., Bloom, B. & Clelland, C. T. Long-term storage of information in DNA. *Science* **293**, 1763 (2001).
8. Yiming, D., Fajia, S., Zhi, P., Qi, O. & Long, Q. DNA storage: Research landscape and future prospects. *Natl. Sci. Rev.* **6**, giz075 (2020).
9. Ceze, L., Nivala, J. & Strauss, K. Molecular digital data storage using DNA. *Nat. Rev. Genet.* **20**, 456–466 (2019).
10. Zhi, P. *et al.* Carbon-based archiving: current progress and future prospects of DNA-based data storage. *GigaScience* **8**, giz075 (2019).
11. Extance, A. How DNA could store all the world's data. *Nature* **537**, 22–24 (2016).
12. Zhirnov, V., Zadegan, R. M., Sandhu, G. S., Church, G. M. & Hughes, W. L. Nucleic acid memory. *Nat. Mater.* **15**, 366–370 (2016).
13. Poltyrev, G. S. Book review. csiszari. and kornerj. "information theory. coding theorems for discrete memoryless systems". *Probl. Peredachi Inf., 1982* 108–111 (1982).
14. Rutten, M. G., Vaandrager, F. W., Elemans, J. A. & Nolte, R. J. Encoding information into polymers. *Nat. Rev. Chem.* **2**, 365–381 (2018).
15. Organick, L. *et al.* Random access in large-scale DNA data storage. *Nat. Biotechnol.* **36**, 242 (2018).
16. Allentoft, M. E. *et al.* The half-life of DNA in bone: measuring decay kinetics in 158 dated fossils. *Proc. R. Soc. B Biol. Sci.* **279**, 4724–4733 (2012).
17. Grass, R. N., Heckel, R., Puddu, M., Paunescu, D. & Stark, W. J. Robust chemical preservation of digital information on DNA in silica with error-correcting codes. *Angewandte Chemie International Edition* **54**, 2552–2555 (2015).
18. KA, W. National human genome research institute. DNA sequencing costs: data from the nhgri genome sequencing program (gsp). *http://www.genome.gov/sequencingcosts* (2020).
19. Church, G. M., Gao, Y. & Kosuri, S. Next-generation digital information storage in DNA. *Science* **337**, 1628 (2012).
20. Goldman, N. *et al.* Towards practical, high-capacity, low-maintenance information storage in synthesized DNA. *Nature* **494**, 77–80 (2013).
21. Yazdi, S. H. T., Yuan, Y., Ma, J., Zhao, H. & Milenkovic, O. A rewritable, random-access DNA-based storage system. *Sci. Rep.* **5**, 14138 (2015).
22. Bornholt, J. *et al.* A DNA-based archival storage system. In *Proceedings of the Twenty-First International Conference on Architectural Support for Programming Languages and Operating Systems*, 637–649 (2016).
23. Erlich, Y. & Zielinski, D. DNA fountain enables a robust and efficient storage architecture. *Science* **355**, 950–954 (2017).
24. Blawat, M. *et al.* Forward error correction for DNA data storage. *Procedia Comput. Sci.* **80**, 1011–1022 (2016).
25. Al-Okaily, A., Almarri, B., Al Yami, S. & Huang, C.-H. Toward a better compression for DNA sequences using huffman encoding. *J. Comput. Biol.* **24**, 280–288 (2017).
26. Reed, I. S. & Solomon, G. Polynomial codes over certain finite fields. *J. Soc. Ind. Appl. Math.* **8**, 300–304 (1960).
27. Rashtchian, C. *et al.* Clustering billions of reads for DNA data storage. In *Advances in Neural Information Processing Systems*, 3360–3371 (2017).
28. Ping, Z. *et al.* Chamaeleo: a robust library for DNA storage coding schemes. *bioRxiv*. 1–14 (2020).
29. Xia, W. *et al.* A comprehensive study of the past, present, and future of data deduplication. *Proc. IEEE* **104**, 1681–1710. https://doi.org/10.1109/JPROC.2016.2571298 (2016).

### Acknowledgements
This work is supported in part by National Key Research and Development Program of China (2020YFA0909100), Shenzhen Key Laboratory of Synthetic Genomics (ZDSYS201802061806209), Guangdong Provincial Key Laboratory of Synthetic Genomics (2019B030301006), and Shenzhen Science and Technology Program (KQTD20180413181837372).

### Author contributions
M.L. drafted and implemented the methods and conducted the mathematical analysis and experiments to validate the ideas. J.W. completed experimental visualisations, conducted experimental analysis, wrote and revised the manuscript, and responded to the reviewers' comments. Y.W. developed the ideas and conceived the study, and also took main responsibility to organise and write the manuscript, responded to the reviewers' comments, and revised the manuscript as well. Y.W., J.D., and Q.J. directed the experiments and commented the results. X.H reviewed the biological aspect of this manuscript. J.D and Q.Q supported the research in both contents and finance. All authors reviewed the manuscript.

## Competing interests
The authors declare no competing interests

## Additional information
**Correspondence** and requests for materials should be addressed to Y.W.

**Reprints and permissions information** is available at www.nature.com/reprints.

**Publisher's note**  Springer Nature remains neutral with regard to jurisdictional claims in published maps and institutional affiliations.

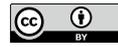 **Open Access**  This article is licensed under a Creative Commons Attribution 4.0 International License, which permits use, sharing, adaptation, distribution and reproduction in any medium or format, as long as you give appropriate credit to the original author(s) and the source, provide a link to the Creative Commons licence, and indicate if changes were made. The images or other third party material in this article are included in the article's Creative Commons licence, unless indicated otherwise in a credit line to the material. If material is not included in the article's Creative Commons licence and your intended use is not permitted by statutory regulation or exceeds the permitted use, you will need to obtain permission directly from the copyright holder. To view a copy of this licence, visit http://creativecommons.org/licenses/by/4.0/.

© The Author(s) 2021